\documentclass[11pt]{article}

\usepackage{amssymb}
\usepackage{amsmath}
\usepackage{graphicx}
\usepackage{epstopdf}
\usepackage[figuresright]{rotating}
\usepackage{verbatim}
\usepackage[section]{placeins}
\usepackage{multirow}
\usepackage{datetime}
\usepackage{cite}
\usepackage[affil-it]{authblk}

\usepackage{setspace}

\newdateformat{mydate}{\monthname[\THEMONTH] \THEDAY, \THEYEAR}
\begin{document}




\title{EXPRESSION OF INTEREST:\\
The Atmospheric Neutrino Neutron Interaction Experiment (ANNIE) }

\date{\mydate\today}


\newcommand{\ANL}{Argonne National Laboratory; Lemont, IL 60439, USA}
\newcommand{\BNL}{Brookhaven National Laboratory;  Upton, NY 11973, USA }
\newcommand{\FNAL}{Fermi National Accelerator Laboratory; Batavia, IL 60510, USA}
\newcommand{\ISU}{Iowa State University; Ames IA 50011, USA}
\newcommand{\NGA}{National Geospatial-Intelligence Agency; Springfield VA 22150, USA}
\newcommand{\OSU}{Ohio State University; Columbus OH 43210, USA}
\newcommand{\Davis}{University of California at Davis; Davis CA 95817, USA}
\newcommand{\Irvine}{University of California at Irvine; Irvine CA 92697, USA}
\newcommand{\UCLA}{University of California at Los Angeles; Los Angeles, CA 90024, USA}
\newcommand{\Chicago}{University of Chicago, Enrico Fermi Institute; Chicago IL 60637, USA}
\newcommand{\Hawaii}{University of Hawaii; Honolulu, HI 96822, USA}
\newcommand{\Queen}{Queen Mary University of London; London E14NS, UK}

\author[1,4]{I.~Anghel}

\author[6]{J.~F.~Beacom}

\author[7]{M.~Bergevin}

\author[4]{G.~Davies}

\author[12]{F.~Di~Lodovico}

\author[10]{A.~Elagin}

\author[10]{H.~Frisch}

\author[10]{R.~Hill}

\author[5]{G.~Jocher}

\author[12]{T.~Katori}

\author[11]{J.~Learned}

\author[10]{R.~Northrop}

\author[10]{C.~Pilcher}

\author[3]{E.~Ramberg}

\author[1,4]{M.C.~Sanchez~\footnote{Corresponding author: Mayly Sanchez (mayly.sanchez@iastate.edu)}}

\author[8]{M.~Smy}

\author[8]{H.~Sobel}

\author[7]{R.~Svoboda}

\author[5]{S.~Usman}

\author[7]{M.~Vagins}

\author[11]{G.~Varner}

\author [1] {R.~Wagner} 

\author[10]{M.~Wetstein~\footnote{Corresponding author: Matthew Wetstein (mwetstein@uchicago.edu)}}

\author[9]{L.~Winslow}

\author[2]{M.~Yeh}

\affil[1]{\ANL}
\affil[2]{\BNL}
\affil[3]{\FNAL}
\affil[4]{\ISU}

\affil[5]{\NGA}
\affil[6]{\OSU}
\affil[7]{\Davis}
\affil[8]{\Irvine}
\affil[9]{\UCLA}

\affil[10]{\Chicago}
\affil[11]{\Hawaii}
\affil[12]{\Queen}


\maketitle



\noindent

\pagebreak

\begin{abstract}
The observation of proton decay (PDK) would rank among the most important discoveries in particle physics to date, confirming a key prediction of Grand Unification Theories and reinforcing the idea that the laws of physics become increasingly symmetric and simple at higher energies. Proposed Water Cherenkov (WCh) detectors, such as Hyper-Kamiokande are within reach of PDK detection according to many general PDK models. However, these experiments will also achieve size scales large enough to see PDK-like backgrounds from atmospheric neutrino interactions at a rate of roughly a few events per year per megaton. Given the rarity of proton decay and significance of the measurement,  the observation of proton-decay should be experimentally unambiguous. 

Neutron tagging in Gadolinium-doped water may play a significant role in reducing these backgrounds from atmospheric neutrinos in next-generation searches. Neutrino interactions typically produce one or more neutrons in the final state, whereas proton decay events rarely produce any. The ability to tag neutrons in the final state provides discrimination between signal and background. Gadolinium salts dissolved in water have high neutron capture cross-sections and produce  $\sim$8 MeV in gammas, several tens of microseconds after the initial event. This delayed 8 MeV signal is much easier to detect than the 2 MeV gammas from neutron capture in pure water. Nonetheless, even the detection of this signature will not be perfectly efficient in large WCh detectors, especially those with low photodetector coverage.

It is not enough to identify the presence or absence of neutrons in an interaction. In proton-decay searches, the presence of neutrons can  be used to remove background events. However, the absence of a tagged neutron is insufficient to attribute confidence to the observation of a proton decay event since the absence of a neutron may be explained by detection inefficiencies in WCh detectors. For moderately efficient neutron tagging and backgrounds peaked at higher neutron multiplicity, the absence of {\it any} neutron would increase confidence in the observation of a PDK candidate event. Calculating an exact confidence for discovery will require a detailed picture of the number of neutrons produced by neutrino interactions in water as a function of momentum transfer. Making this measurement in a neutrino test-beam is thus critical to future proton-decay searches. 

The neutron tagging techniques based on such measurement will also be useful to a broader program of physics beyond proton decay. For example, in the detection of diffuse supernova neutrino background, neutron tagging can be used to separate between genuine neutrinos and various radiogenic and spallation backgrounds. In the event of a core collapse supernova, the detection of neutrons can be used to help discriminate among different interactions in the water such as inverse beta decay and neutrino-electron scattering.

In this white paper we propose the Atmospheric Neutrino Neutron Interaction Experiment (ANNIE), designed to measure the neutron yield of atmospheric neutrino interactions in gadolinium-doped water. While existing experiments such as Super-Kamiokande have attempted {\it in situ} measurements of neutron yield, the analyses were limited by detection inefficiencies and unknowns in the flux and energy of atmospheric neutrinos. ANNIE would represent a small, dedicated experiment designed to make this measurement on a beamline with known characteristics.

An innovative aspect of the ANNIE design is the use of precision timing to localize interaction vertices in the small fiducial volume of the detector. We propose to achieve this by using early prototypes of LAPPDs (Large Area Picosecond Photodetectors). This experiment will be a first application of these devices demonstrating their feasibility for WCh neutrino detectors. The ideas explored by ANNIE could have a transformative impact on water Cherenkov, scintillation, and other photodetection-based neutrino detector technologies.

\end{abstract}

\newpage
\tableofcontents
\setcounter{tocdepth}{3}
\newpage

\section{Motivation for the Measurement of Neutron Yield in Water}
\label{sec-physics}

The ability to detect final state neutrons from nuclear interactions would have a transformative impact on a wide variety of physics measurements in very large Water Cherenkov (WCh) and liquid scintillator (LS) detectors~\cite{FSneutrons}.  Neutrino interactions in water typically produce one or more neutrons in the final-state. Tagging events by the presence and number of final-state neutrons can provide physics analyses with a better handle for signal-background separation and even allow for more subtle discrimination between different varieties of neutrino interactions. For example, the main background on proton decay experiments arise from atmospheric neutrino interactions. These interactions almost always produce at least one final-state neutron, whereas proton decays are expected to produce neutrons less than 10$\%$ of the time~\cite{Ejiri}.

A promising technique for detecting final state neutrons is the search for a delayed signal from their capture on Gadolinium dissolved in the target liquid. Even moderately energetic neutrons ranging from tens to hundreds of MeV will quickly lose energy by collisions with free protons and oxygen nuclei in water. Once thermalized, the neutrons are captured, creating unstable nuclei and excited states that emit radiation. Neutron capture in pure water typically produces around 2.2 MeV in gamma particles ($\gamma$)~\cite{Ncapturewater}. However, these low energy photons produce very little optical light and are difficult to detect in large WCh tanks. The introduction of Gadolinium (Gd) salts dissolved in the target liquid is proposed as an effective way to improve the detection efficiency of thermal neutrons. With a significantly larger capture cross-section (49,000 barns compared with 0.3 barns on a free proton), Gd-captures happen roughly 10 times faster, on the order of tens of microseconds~\cite{NcaptureGd}. In addition, the Gd-capture produces an 8 MeV cascade of typically 2-3 gammas, producing sufficient optical light to be more reliably detected in large volumes.

A major limitation on the effective execution of neutron tagging techniques comes from large uncertainties on the nuclear mechanisms that produce neutrons and consequently on how many neutrons are produced by high energy (GeV-scale) neutrino interactions. The number of neutrons is expected to depend on the type of neutrino interaction and on the momentum transfer with higher energy interactions producing more than one neutron. However, the exact number of neutrons is determined by a variety of poorly understood nuclear processes and therefore not well-known. 

It is not enough to identify the presence or absence of neutrons in an interaction.  While the presence of neutrons can  be used to remove background events, the absence of a tagged neutron is insufficient to attribute confidence to the discovery of a proton decay observation. The absence of a neutron may be explained by detection inefficiencies in the WCh detector. On the other hand, if typical backgrounds consistently produce {\it more} than one neutron, the absence of {\it any} neutron would increase the confidence in a PDK-like event. Calculating an exact confidence for discovery will require a detailed picture of the number of neutrons produced by neutrino interactions in water as a function of momentum transferred. 

\begin{figure}
	\begin{center}
		\includegraphics[width=0.5\linewidth]{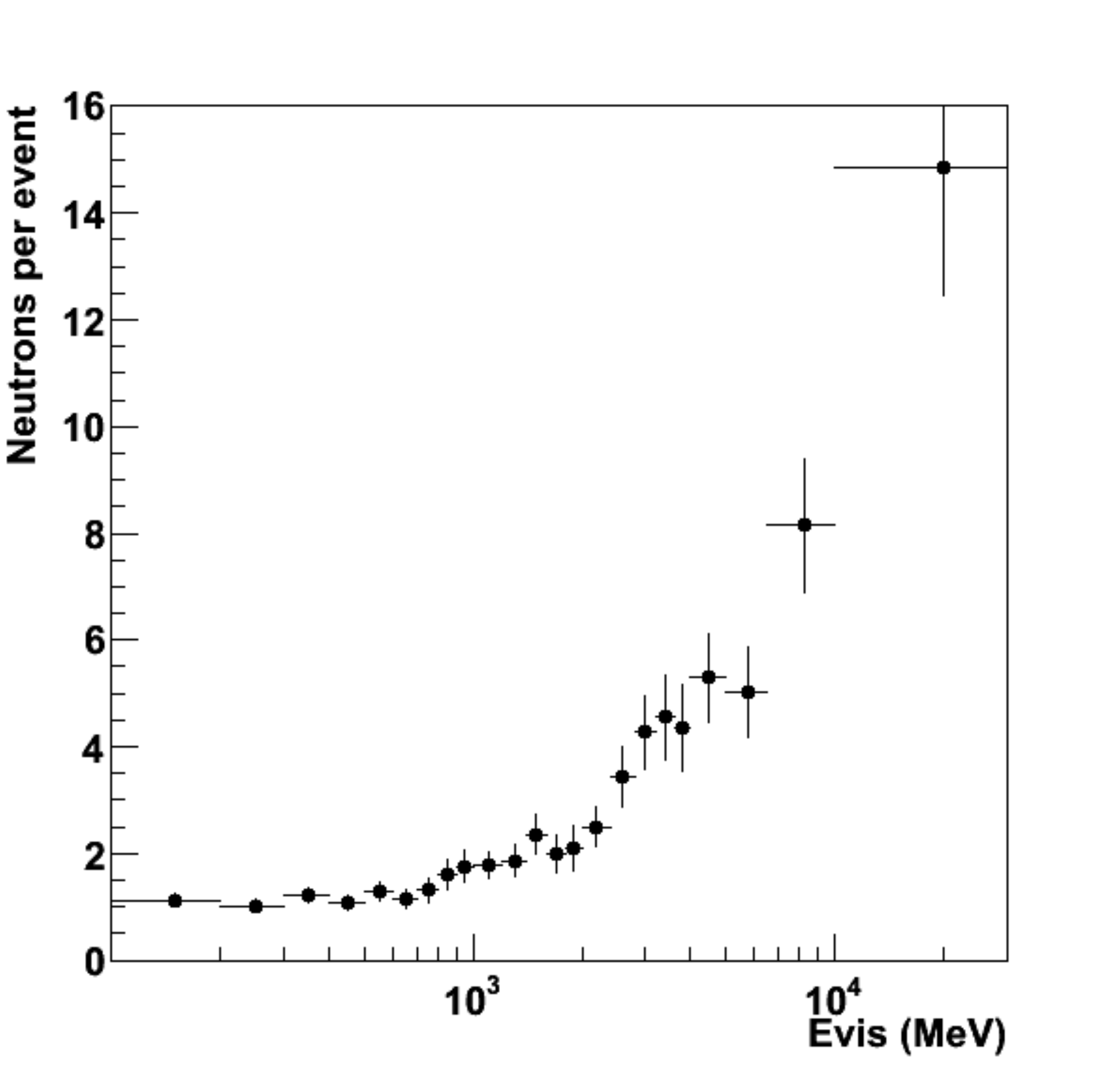}
	\end{center}
	\caption{Measurement of neutron multiplicity in pure water versus visible energy by the Super-K collaboration~\cite{SKneutronyield}.}
	\label{SKneutronmult}
\end{figure}

The Super-Kamiokande (Super-K) collaboration has measured the final state neutron abundance. Fig~\ref{SKneutronmult} shows the neutron multiplicity as a function of visible energy from atmospheric neutrino interactions in water, as detected by the 2.2~MeV capture gamma in Super-K~\cite{SKneutronyield}. However, the Super-K analysis is limited by uncertainties on the detection efficiencies for the 2.2~MeV gammas and on the flux of atmospheric neutrinos. Additionally, neither the neutrino energy nor the momentum transfer to nucleus can be measured precisely. Therefore, it is difficult to incorporate these data into background predictions for proton decay, although it is encouraging that the multiplicity at $\sim$1 GeV seems to be close to two. 

Therefore, there is a clear need for a dedicated measurement of neutron yield. Such detailed measurement of the neutron multiplicity is possible in a beam with atmospheric neutrino-like energy spectrum. We propose to build such an experiment. The Atmospheric Neutrino Neutron Interaction Experiment would consist of a small, economical Water Cherenkov detector deployable on the intense Booster Neutrino Beam at Fermilab, and would largely rely on existing infrastructure. The main deliverable from this experiment is a measurement of the final-state neutron abundance as a function of momentum transfer from charged current (CC) neutrino interactions. This measurement is similar to the plot shown in Fig~\ref{SKneutronmult}, except that we would reconstruct the total momentum transfer rather than visible energy and our detector will be optimized for efficient detection of captured neutrons produced in the fiducial volume. Furthermore, it may be possible to separate the data between a variety of CC event types and possibly neutral current (NC) interactions. These data will provide an essential input to PDK and neutrino-interaction Monte Carlo models to aid in calculating detection efficiencies, expected background rates, accurate limits, and confidence levels. They can also be used to better constrain nuclear models of neutrino interaction physics and are therefore interesting in themselves.

\section{Potential Physics Impact}

\subsection{Understanding a Critical Background in Proton Decay}

One of the ``Big Ideas" in particle physics is the notion that at higher energies, the laws of physics become increasingly symmetric and simple. In the late 1970s it was suggested that, barring perturbations from other processes, the three running coupling constants become similar in strength in the range of $10^{13}-10^{16}$ GeV~\cite{RabyNakamura}. This convergence hints that the electromagnetic, weak, and strong forces may actually be a single force with the differences at low energy being due to the details of the exchange particle properties and the resulting vacuum polarization. This so-called ``Grand Unified Theory" (GUT) is a touchstone of particle physics in the late 20th and early 21st centuries. Theories ranging from supersymmetry (SUSY) to a wide class of string theories all have this basic ``Big Idea". A major challenge for experimental particle physics is how to determine if it is really true.

A convergence of the coupling constants at a very high ``unification energy" implies that there may be a single force that could connect quarks and leptons at that scale. Such reactions would violate baryon (B) and lepton (L) number by the exchange of very heavy bosons with masses in the range of the unification energy scale. Since that scale is far beyond the reach of any conceivable accelerator, they would only manifest themselves at our low energy scale via virtual particle exchange leading to rare reactions that would violate B and L. This would mean that normal matter (e.g., protons, either free or in nuclei) would not be stable but would decay with some very long lifetime. This phenomenon, generically called {\it proton decay} although neutrons in nuclei are included, has been searched for in a series of experiments dating back more than thirty years. Its discovery would be nothing short of revolutionary.

Proton decay final states depend on the details of a given theory. Experimentally, the modes $p \rightarrow e + \pi_0$ and $p \rightarrow K^+ + \nu$ are common benchmarks. The former represents the lightest anti-lepton plus meson final state, typical for the case where the first generation of quarks and leptons are grouped in a single multiplet, as in SU(5). The second is typical of supersymmetric grand unified theories where dimension-5 operators induce decays that span generations, hence requiring a strange quark. Current limits from SK for these two modes are $8.2\times 10^{33}$ and $2\times10^{32}$ years, respectively~\cite{{SuperK2},{SuperK3}}.

\subsubsection{$p \rightarrow e + \pi_0$}

\begin{figure}
	\begin{center}
		\includegraphics[width=0.6\linewidth]{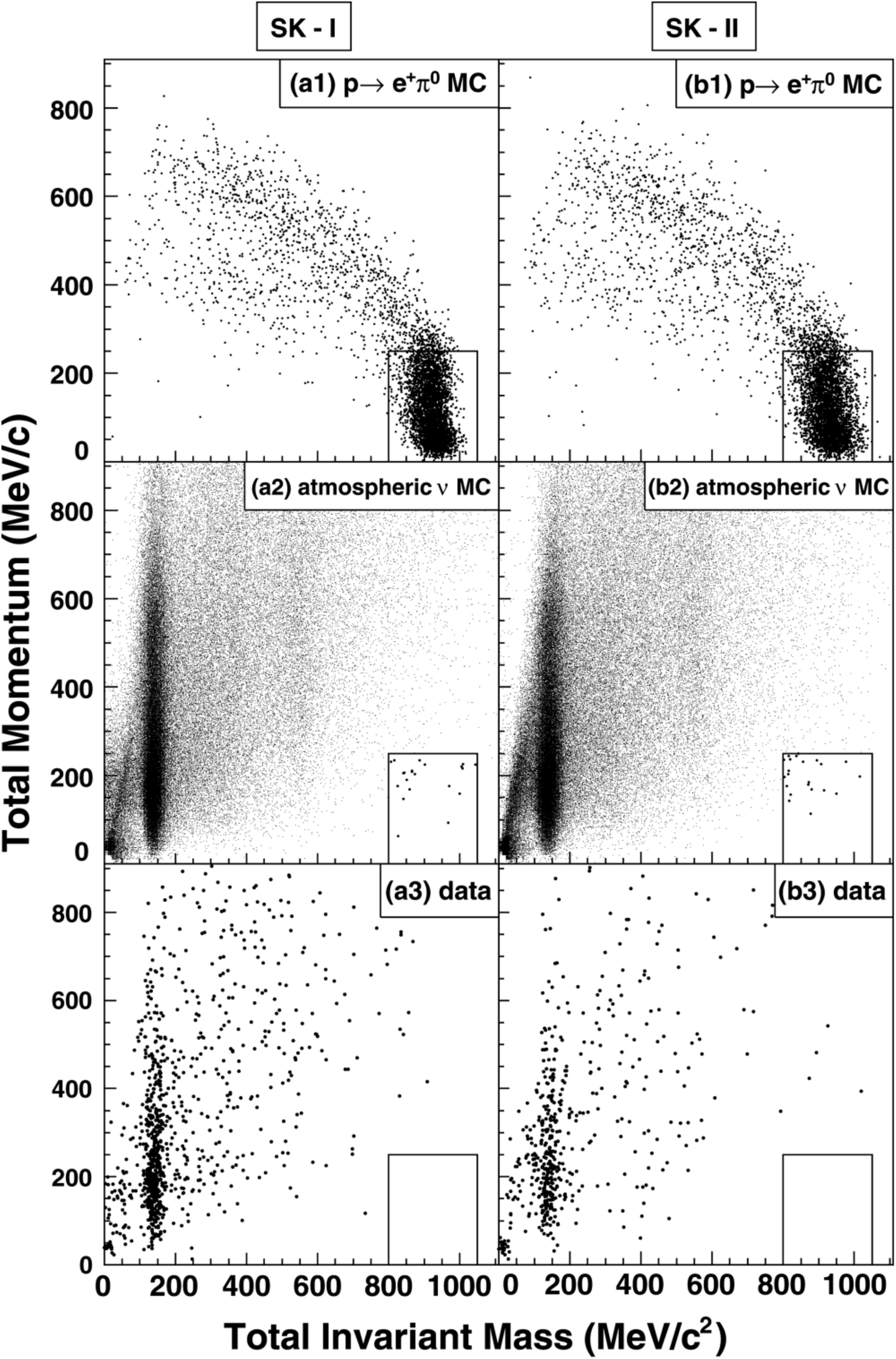}
	\end{center}
	\caption{The reconstructed kinematics of proton decay events in Super-K Monte Carlo (a1,b1), compared with those of atmospheric neutrino Monte Carlo (a2,b2) and data (a3,b3). Atmospheric neutrino events that fall in the signal region of (a2,b2) are enlarged (Ref~\cite{SuperK2}).}
	\label{pdkbkgd}
\end{figure}

It is instructive to describe the analysis currently being used by the Super-K experiment. This analysis consists of (1) selection of events in the detector that have three showering tracks, (2) a requirement that at least one combination of tracks gives an invariant mass close to that of the $\pi^0$ (85-185 MeV), (3) a requirement that there was no follow-on Michel electron (indicating that there was a muon in the event), and (4) that the invariant mass be near that of the proton (800-1050 MeV) and the unbalanced momentum be less than 250 MeV/c. Figure~\ref{pdkbkgd} (reproduced from ~\cite{SuperK2}) shows the invariant mass-unbalanced momentum distributions for two versions of Super-K (the left plots have twice the number of PMTs as the right plots) for the proton decay MC (top), atmospheric neutrino background MC (middle), and data (bottom). At 0.141 Mton-years there are no candidates.

The selection efficiency of the Super-K analysis was estimated to be $45\%$, with an uncertainty of $19\%$ dominated by nuclear effects (mainly pion interactions in the oxygen nucleus). In the center plots, the incursion of background events into the signal region is clearly seen. The MC gives a background estimate of $2.1 \pm 0.9$ events/Mton-year, which is consistent with direct measurements made in the K2K 1-kton near detector ($1.63_{-0.33}^{+0.42}$(stat)$_{-0.51}^{+0.45}$(syst) events/Mton-year)~\cite{SuperK3}. 

According to the Super-K MC, about $81\%$ of the background events are CC, with $47\%$ being events with one or more pions, and $28\%$ being quasi-elastic. In many cases, a $\pi^0$ is produced by an energetic proton scattering in the water. These events could be rejected by means other than invariant mass and unbalanced momentum. Neutron tagging has been proposed as a key method for doing this. Many of these background-producing events should be accompanied by one or more neutrons in the final state. This is because to look like a proton decay, there needs to be significant hadronic activity in the event, and there are many ways to produce secondary neutrons:

\begin{itemize}
\item direct interaction of an anti-neutrino on a proton, converting it into a neutron
\item secondary (p,n) scattering of struck nucleons within the nucleus
\item charge exchange reactions of energetic hadrons in the nucleus (e.g., $\pi^- + p \rightarrow n + \pi^0$)
\item de-excitation by neutron emission of the excited daughter nucleus
\item capture of $\pi^-$ events by protons in the water, or by oxygen nuclei, followed by nuclear breakup
\item secondary neutron production by proton scattering in water
\end{itemize}


Unfortunately, simulations of these processes are not currently data-driven. It is thus not possible to reliably predict the number of neutrons produced following a neutrino interaction. There is some experimental data from Super-K that supports the idea that atmospheric neutrino interactions in general have accompanying neutrons. 

This is to be contrasted with signal proton decay events, which are expected to produce very few secondary neutrons. Using general arguments, it is expected that more than $80\%$ of all proton decays should {\it not} have an accompanying neutron: 
\begin{itemize}
\item For water, $20\%$ of all protons are essentially free. If these decay, there is no neutron produced as the $\pi^0$ would decay before scattering in the water, and 400 MeV electrons rarely make hadronic showers that result in free neutrons.
\item Oxygen is a doubly-magic light nucleus, and hence one can use a shell model description with some degree of confidence. Since two protons are therefore in the $p_{1/2}$ valence shell, if they decay to $^{15}N$, the resultant nucleus is bound and no neutron emission occurs except by any final state interactions (FSI) inside the nucleus. 
\item Similarly, if one of the four protons in the $p_{3/2}$ state decays, a proton drops down from the $p_{1/2}$ state emitting a 6 MeV gamma ray, but the nucleus does not break up except by FSI.
\item Finally, if one of the two $s_{1/2}$ protons decays, there {\it is} a chance that the nucleus will de-excite by emission of a neutron from one of the higher shells.
\end{itemize}

Detailed nuclear calculations by Ejiri~\cite{Ejiri} predict that only $8\%$ of proton decays in oxygen will result in neutron emission. This means that only approximately $6\%$ ($8\%$ of $80\%$) of all proton decays in water should result in neutrons (ignoring FSI production by proton decay daughters). Therefore neutron tagging to reject atmospheric neutrino backgrounds incurs in a modest loss of signal efficiency. In this proposal our goal is to measure the neutron yield in neutrino interactions as a function of momentum transfer. This will allow us to assess the effectiveness of such strategy.

\begin{figure}
	\begin{center}
		\includegraphics[width=0.5\linewidth]{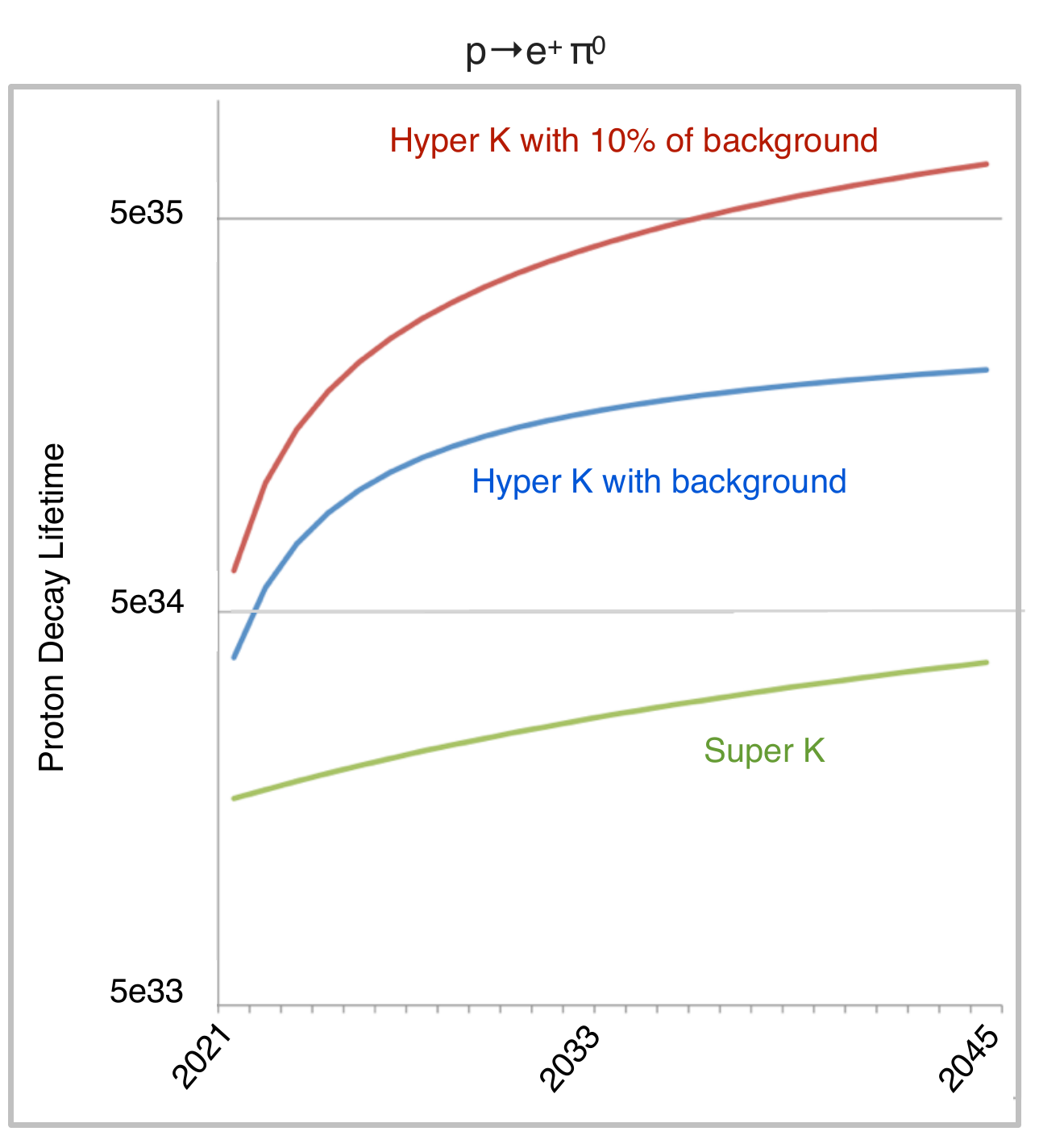}
	\end{center}
	\caption{Proton decay sensitivity in the $p \rightarrow e + \pi_0$ channel at Super-K and a 0.5 Mton detector with two different background assumptions.}
	\label{bkgdlimits}
\end{figure}

As an illustration, Figure~\ref{bkgdlimits} shows the sensitivity of Super-K (green) if it continues to run another 35 years, assuming the expected background rate from atmospheric neutrinos remains unchanged. Uncertainties in the background spectrum are taken into account, and the curves shown are $90\%$ c.l. limits.  Also shown (in blue) is the sensitivity of a 0.5-Mton detector with similar background estimations as Super-K running for a similar amount of time. Substantial background reduction using neutron tagging techniques is expected to significantly improve the sensitivity and discovery potential of very large WCh detectors. For example, the curve shown in red highlights the impact of such background reduction. However, the precise background rejection efficiencies have not been demonstrated. ANNIE will accurately evaluate and demonstrate the potential of this method.

\subsubsection{$p \rightarrow K^+ + \nu$ }

As another example, for the $p \rightarrow K^+ + \nu$ mode, the $K^+$ is below the Cherenkov threshold, requiring a search for the decay of a kaon at rest. There is significant atmospheric neutrino background in the dominant ($63\%$) decay mode of $K^+ \rightarrow \mu + \nu_{\mu}$. Super-K uses the prompt nuclear de-excitation gamma ray (6.3 MeV) from the residual $^{15}N$ nucleus to reject background events. Analysis of the hadronic mode, $K^+ \rightarrow \pi^0 \pi^+$ ($21\%$), is hampered by the fact that $\beta_{\mu+}$ = 0.87, so that the amount of Cherenkov light emitted by the decay muon (from the $\pi^+$) is near the detectable threshold. Expectations are that background events will be seen in this mode at a rate of ~8 events/Mton-year. The combined efficiency for the prompt gamma tag of $K^+ \rightarrow \mu + \nu_{\mu}$ plus $K^+ \rightarrow \pi^0 \pi^+$ is $14\% \pm 2\%$ with an expected background of $1.2 \pm 0.4$ events/100-kton/year. Thus even though Super-K does not currently have a candidate, it is expected that this mode will soon start generating background. If a significant fraction of these events could be rejected, sensitivity would continue to rise relatively linearly in a very large detector.

\subsection{Neutron Tagging to Improve Identification of Supernova Neutrino Interactions}

Supernova explosions throughout the universe left behind a diffuse supernova background of neutrinos that may be detected on Earth. The flux and spectrum of this background contains information about the rate of supernova explosions as well as their average neutrino temperature.  The main detection channel for supernova relic neutrinos in water Cherenkov detectors comes from positrons emitted by inverse $\beta$ decay reactions. Above $\sim 20$ MeV, the dominant background is due to the decay of sub-Cherenkov threshold muons from atmospheric neutrino interactions. This could be greatly reduced by tagging the neutron that accompanies each inverse $\beta$ reaction. A 200-kton detector loaded with gadolinium and at sufficient depth may be within reach of detecting this neutrino flux~\cite{LBNESN}. In order to achieve this, understanding neutron yields can be used to help statistically discriminate among various radiogenic, spallation and neutrino backgrounds.  

A nearby core collapse supernova will provide a wealth of information via its neutrino signal. The neutrinos are emitted in a burst of a few tens of seconds duration, with about half in the first second. Energies are in the few tens of MeV range, and the luminosity is divided roughly equally among flavors. Neutrino densities in the core are so high that neutrino-neutrino scattering plays a significant role in the dynamics, leading to non-linear oscillation patterns, highly sensitive to fundamental neutrino properties and even new physics. Accurate measurements of the energies, flavors, and time dependent fluxes would also allow one to set limits on coupling to axions, large extra dimensions, and other exotic physics~\cite{Raffelt}. From these details, one could also learn about the explosion mechanism, accretion, neutron star cooling, and possible transitions to quark matter or to a black hole. Neutron tagging would be essential in building a more complete picture of the SN neutrino flux, helping to more efficiently identify neutral current interactions, and separate neutrino-electron scattering in the water, which do not produce any neutrons. 

\subsection{Testing Nuclear Models of Neutrino Interactions}

There is a growing interest among the neutrino cross-section community in better understanding nuclear effects on neutrino interactions. Most of the current and future long-baseline neutrino oscillation experiments are designed to measure neutrinos with energies below 10 GeV. Nuclear effects play a significant role in this regime, as demonstrated by the recent T2K oscillation result~\cite{T2K2}, where the neutrino interaction model is the largest systematic error.

The MiniBooNE experiment has published a first double differential cross section for CC quasi-elastic (QE) interactions~\cite{MB1,MB2}. Many aspects of this precision measurement are not understood by traditional nuclear models based on the impulse approximation~\cite{Benhar}. The MiniBooNE data may be better described by models including two-body currents, where low-energy neutrinos scatter off of correlated pairs of nucleons~\cite{Martini,Nieves}. Confirming such processes and incorporating them into oscillation analyses is now a major goal of the cross-section community. A predicted consequence of two-body currents is a higher multiplicity of final-state nucleons~\cite{Martini}. An experiment, like ANNIE, with neutron tagging abilities would provide a unique opportunity to study some of these effects.

Final-state neutrons can also be used used in the statistical separation between NC interactions and CC interactions. In neutrino-mode, neutron multiplicity is expected to be lower for CC interactions. This feature can be used to distinguish $\nu_e$ oscillation candidates from NC backgrounds, such as $\pi^0$ or photon production~\cite{MBp}. ANNIE is in a position to study the feasibility of this technique in water.

\section{The ANNIE Experiment} 
\label{sec-experiment}

\begin{figure}
	\begin{center}
		\includegraphics[width=0.45 \linewidth]{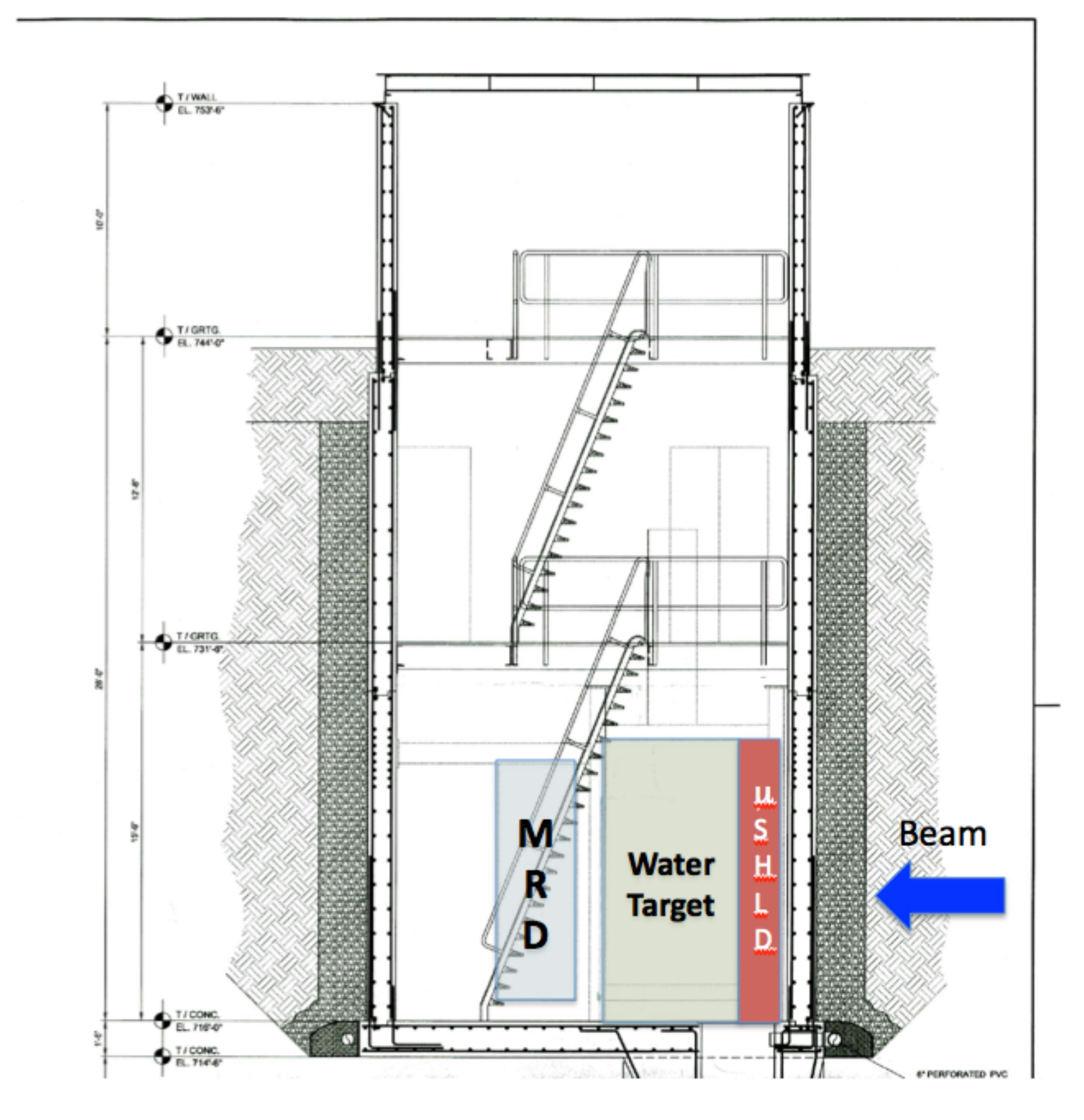}
	\end{center}
	\caption{ANNIE in the SciBooNE Hall.}
	\label{anniehall}
\end{figure}

We propose making a systematic measurement of the neutron yield from neutrino interactions of energies similar to atmospheric neutrinos. We will do this by utilizing the existing ``SciBooNE" hall in the FNAL booster beam (Fig~\ref{anniehall}). The hall is currently unused, and contains an iron-scintillator sandwich detector that was used to range out and fit the direction of daughter muons from neutrino interactions in the SciBooNE target~\cite{Sciboone}. This detector, called the Muon Range Detector (MRD) is available and needs only to be connected to HV and a suitable DAQ system. We plan to put a Gd-doped water target in front of the MRD that is doped with gadolinium and sufficiently instrumented to be able to stop and detect the capture gammas from primary and secondary neutrons. We have named this beam test the Atmospheric Neutrino Neutron Interaction Experiment, or ANNIE.

\begin{figure}
	\begin{center}
		\includegraphics[width=0.60 \linewidth]{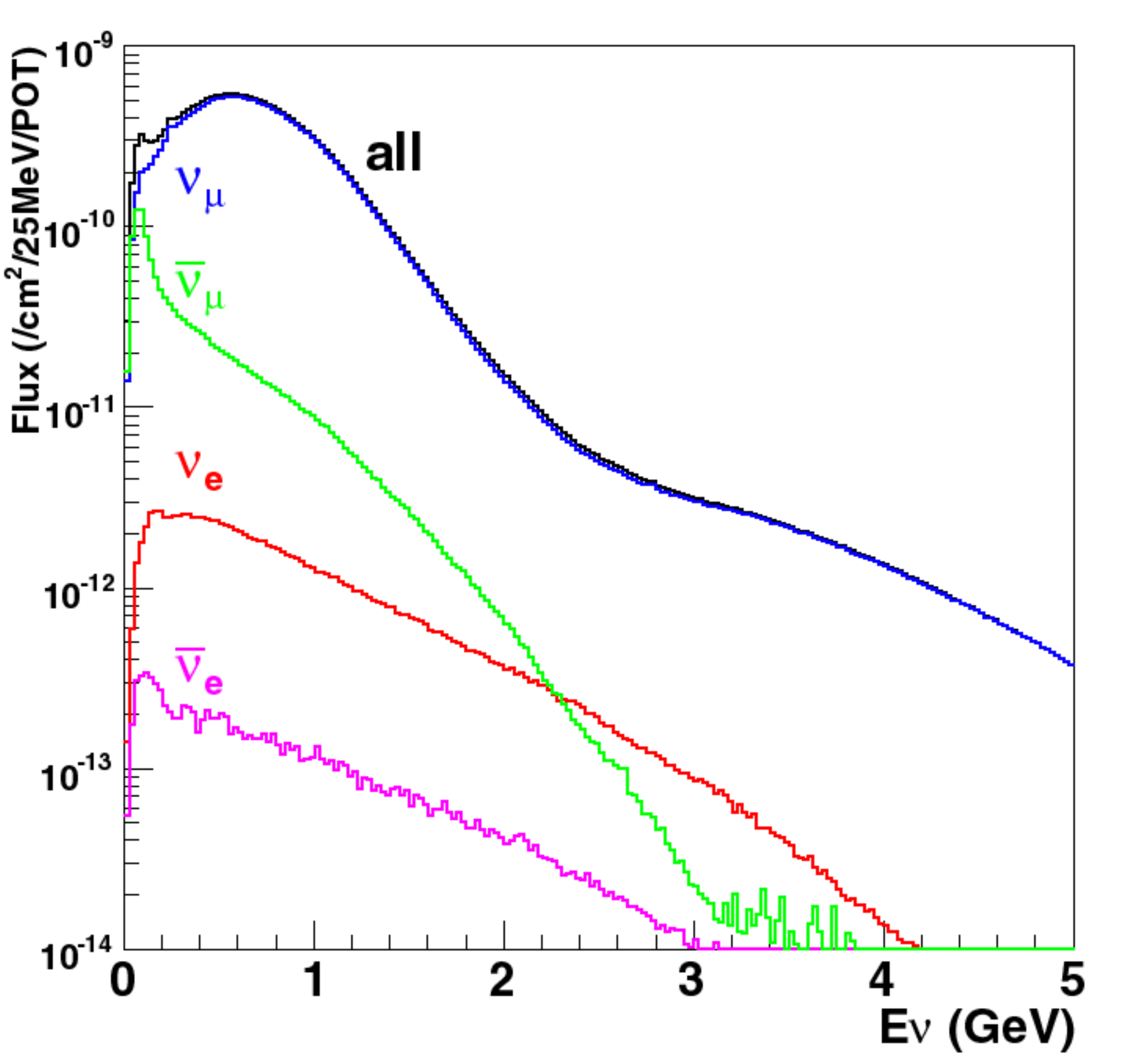}
	\end{center}
	\caption{Fluxes expected in ANNIE Hall.}
	\label{anniefluxes}
\end{figure}

The sequence of the measurement is as follows: (1) The booster beam runs at 15 Hz, with roughly 4x$10^{12}$ protons-on-target (POT) per spill. These are delivered in 81 bunches over a $1.6$ $\mu$sec spill time to a target and horn combination 100 m upstream of the SciBooNE hall. The cycle time is about 2 seconds. This beam is about $93\%$ pure $\nu_{\mu}$ (when running in neutrino mode) and has a spectrum that peaks at about 700 MeV (Fig.~\ref{anniefluxes}). (2) A neutrino interaction in the water target produces a flash of light with no signal in the front anti-coincidence counter. (3) A selection is made on only those events that are ``CCQE-like", i.e., there is a single muon track in the MRD that points back to the rough position of the vertex in the target. (4) Following a valid CCQE candidate, a gate is opened to count neutron capture events in the target for about 100 $\mu$s, or about three capture times. If the vertex is restricted to the central volume of the water target, then there should be several hadron scattering lengths in all directions, which should be enough to slow down and stop neutrons in the range of 110 MeV. Higher energy neutrons could require external counters, which is not being proposed here but could be part of a future upgrade. By measuring the muon direction to a precision of roughly $10^{\circ}$ and the muon momentum (from a range measurement) to roughly $20\%$, we will be able to accurately reconstruct the multiplicity as a function of the momentum transfer to the nucleus from the neutrino. This is desirable to facilitate incorporation into an atmospheric neutrino MC. 

Studies of the optimization of the water target and MRD cuts would be made during the first year of the proposal. It is expected that almost all the equipment for ANNIE will be recycled from other experiments, and this will be noted in the detailed description of the components. 

Table~\ref{expectedrates} gives the number of interactions per ton per $10^{20}$ POT at the ANNIE location. Note that several $10^{20}$ POT is to be expected during the next neutrino run starting at the beginning of 2014 and going for several years to allow MicroBooNE to run. The major components of the ANNIE detector are described below.

\begin{table*}
  \begin{center}
    \begin{tabular} {| l | c | c | c |}
     \hline
     $\nu$-type & Total Interactions & Charged Current & Neutral Current\\
     \hline
     \hline
     $\nu_{\mu}$ & 10210 &7265 & 2945\\
     \hline
     $\bar{\nu}_{\mu}$ & 133 &88 & 45\\
     \hline
     $\nu_{e}$ & 70 & 52 & 20\\
     \hline
     $\bar{\nu}_{e}$ & 4.4 &3 & 1.4\\
     \hline
    \end{tabular}
    \caption{Rates expected in 1 ton of water with 1x$10^{20}$ POT exposure at ANNIE Hall.}
  \end{center}
  \label{expectedrates}
\end{table*}


\subsection{Front Anti-Coincidence Counter (FACC)}

Since we expect roughly 50 events/day/ton spread over roughly 40,000 cycles, it is clear that only about $0.2\%$ of spills will have a neutrino event if we utilize only the central part of the target. It will be important to reject the muons that come from neutrino-induced events in the rock, which will be much more frequent. Hence, an efficient muon rejection counter is needed in front of the water target. There are many potential sources of such counters from previous experiments, including counters available from Iowa State. We will test and decide on the exact counter configuration based on simulations in the first year. 

\subsection{Muon Range Detector (MRD)}

This consists of 12 slabs of 2-inch-thick iron plates sandwiched between 13 alternating vertical and horizontal layers of 20-cm-wide, 0.6-cm-thick plastic scintillation panels read out by 362 two-inch PMTs. The vertical layers are arranged in a 2x15 pattern of 138-cm-long paddles, while the horizontal layers are in a 2x13 array of 155-cm-long paddles. Currently, this device is sitting unused in the SciBooNE Hall and needs only some minor repairs plus HV and DAQ readout systems. These systems could be provided by FNAL and/or the UK collaborators.

\subsection{Water Target}

The footprint for the water target is essentially that of the SciBooNE detector, a volume roughly 3 m x 3 m and 1.9 m thick. It may also be possible to enlarge the detector along the beam direction by moving the MRD one meter further back. The current plan is to contain the target volume in a single water tank made of steel with a teflon liner and open top covered by black plastic. The target will be instrumented by ~80 eight-inch PMTs on top looking down and 20 ten-inch PMTs on the bottom looking up. The top PMTs will be supplied by UC Irvine from Super-K spares, and the bottom PMTs will be from LLNL or UC Davis from existing stock. These PMTs are already water-potted, though the UC Irvine and UC Davis PMTs may need some refurbishment and testing. The water system will be provided by an existing portable RO system from UC Davis, and a gadolinium recirculation system (if needed) will be supplied by Irvine. The current concept for localizing the vertex inside the target is through timing based reconstruction, as will be discussed below.

\subsection{Timing Based Fiducial Cuts Using LAPPDs}

In order to select events away from the detector wall, we propose to use vertex reconstruction based on the arrival time of emitted light. This is the simplest option, requiring neither segmentation of the already small target nor the introduction of new materials with unknown neutron capture properties. Given the few-meter length scale of the detector, timing based reconstruction is a challenge. Typical drift times for direct light are below 10 nanoseconds, so it is unlikely that conventional PMTs, with few-nanosecond time resolutions will be good enough to localize the vertex. Rather, we intend to use early commercial prototypes of Large Area Picosecond Photodetectors (LAPPDs) with single photoelectron time resolutions below 100 picoseconds. These will be placed primarily on the surface facing the beam, though some might be distributed on the other walls. 

The interaction point can reconstructed using several different approaches, thus providing an effective handle on the location of the interaction point. The ideal method is to fit for the earliest light emitted along the muon track, using the track parameters extracted from the muon range detector. It may be possible to use the isotropic light emitted from the neutron captures themselves to determine where in the volume the captures are happening. While basic simulations of the ANNIE detector already exist, we will build a fully integrated Geant4 simulation of the experiment over the next several months. Through these simulations, we can address the technical design issues required for the success of this effort.

\section{R$\&$D Tasks}
\label{sec-rd}

\subsection{Operation of LAPPDs in a Water Cherenkov Detector}

The Large Area Picosecond Photodetector (LAPPD) project was formed to develop new fabrication techniques for making large-area (8" x 8") MCP photodetectors using low-cost materials, well established industrial batch techniques, and advances in material science~\cite{LAPPD}. 

LAPPDs may be a transformative technology for WCh detectors. While conventional photomultipliers are single-pixel detectors, LAPPDs are imaging tubes, able to resolve the position and time of single incident photons within an individual sensor. This maximizes use of fiducial volume as it allows for reconstruction of events very close to the wall of the detector, where the light can only spread over a small area. The simultaneous time and spatial resolutions of the LAPPDs, at better than 100 picoseconds and 3mm for single photons, represent a major improvement over conventional PMTs. Preliminary Monte Carlo studies indicate that the measurement of Cherenkov photon arrival space-time points with resolutions of 1 cm and 100 psec will allow the detector to function as a tracking detector, with track and vertex reconstruction approaching size scales of just a few centimeters~\cite{fastneutrino}. Imaging detectors would enable photon counting by separating between the space and time coordinates of the individual hits, rather than simply using the total charge. This means truly digital photon counting and would translate directly into better energy resolution and better discrimination between dark noise and photons from neutron captures. Finally, at a thickness of less than 1.5 cm, LAPPDs maximize the use of the limited fiducial volume available to small detectors. 

As the LAPPD effort transitions into the next stage of the project, more devices may become available through the commercialization process. The DOE has already awarded a Phase I SBIR to Incom Inc to begin the commercialization process, with $\$3$M possible for Phase II. An industry standard is around 3-4 years as the expected timeframe for commercialization. It is expected that some small numbers of LAPPDs will be available sooner than that. Such time scales  work well with the planned schedule for ANNIE.

Some application-specific detector development work will be necessary to ready LAPPDs for use in WCh detectors. In this section we discuss some of the steps necessary to ready these sensors for use in a WCh.

\subsubsection{Adaptation of Full Waveform Digitizing Front-End Electronics and DAQ}

Groups at the University of Chicago and University of Hawaii have already made and tested complete front-end systems for LAPPDs, built around a new class of low power, custom designed waveform sampling chips~\cite{{UCelectronics2},{UHawaiielectronics},{UHawaiielectronics2},{UHawaiielectronics3},{UHawaiielectronics4}}. We would adapt these systems for use in our Water Cherenkov Detector. Some development work will be necessary to address the unique needs of the ANNIE detector. ANNIE will require an integrated system with both LAPPDs and conventional PMTs. Because of the near-surface operation and intense light from signal events spread over the small area of the detector walls, the front-end and DAQ systems will be designed to handle pileup.

\subsubsection{Operational demonstration of LAPPD systems in small test facilities}

Once commercially available, early LAPPD prototypes will be thoroughly tested, first on a test bench and then in a scaled-down operational context. A large variety of testing facilities and fixtures are available for the task. In addition, there is a broader community of interested ``early adopters" with plans to use LAPPDs in similar experimental contexts. A facility at the Advanced Photon Source at ANL is designed to characterize the time-response of LAPPD detectors using a fast-pulsed laser~\cite{APS}. Fixtures exist to test complete, resealable LAPPD systems with electronics. These facilities can be used to benchmark the key resolutions of detector systems and study the effects of pileup from coincident pulses on the same anode strip. Moreover, these tests can begin even before the first sealed commercial prototypes become available. Eric Oberla at University of Chicago is building a small, tubular WCh detector for timing-based tracking of cosmic muons using commercially available Planacon MCPs. The electronics are designed using the LAPPD front-end system, which will provide an excellent opportunity for developing operational experience and testing aspects of the front-end design. It may be possible to deploy this detector on a Fermilab test-beam. Finally, there are discussions of performing tests either with a radioactive calibration sample or a test beam on a small target mass coupled to a handful of LAPPDs. 

\subsubsection{Submersion of LAPPDs in water}

High voltage connections for LAPPDs will be made on the front window. Work will be carried out to design and test a scheme to make these connections and operate the front-end electronics in water. In addition to answering these specific technical challenges, the ANNIE effort will also provide critical feedback on the performance and long-term operation of LAPPDs, thus helping to expedite the commercialization process. 

\subsection{Implementation of Event Reconstruction Strategy} 

In parallel to the development of the hardware, there is an existing effort to develop the simulations and reconstruction work necessary to guide the design of the detector and make full use of the LAPPDs. Groups at Iowa State, University of Chicago, Queen Mary University, and UC Irvine will significantly contribute to this effort. The groups will use a variety of existing resources and their previous experience in this area. 

\subsubsection{Simulations-Guided Optimization of Detector Design}

Critical to the ANNIE measurement is the design of a detector large enough to ensure capture of neutrons originating from a modest fiducial volume, and the efficient detection of the captures. Given the known physics of neutron scattering and capture in Gd-doped water, we will use Geant simulations to answer these questions and thereby set the dimensions and determine the needed photocathode coverage of our detector. Simulations will also be needed to answer questions regarding possible backgrounds from cosmic rays, neutrino interactions from the rock behind the wall of the detector hall, as well as cosmogenic backgrounds. Determining the rates and light yield of this pileup is necessary for the design of the trigger system and the LAPPD readout system. Also, light from signal events may be concentrated over small areas and thus the pileup of hits on individual anode channels must be understood. Optimizing pattern matching algorithms, and determining appropriate buffer sizes and tolerable dead times will also be important.

Production of LAPPDs at Incom will occur in small test batches, with volume increasing in time. Consequently, ANNIE will likely start with sparse LAPPD coverage: roughly five 20cm x 20cm LAPPDs to cover the entire 3m x 3m wall facing the beam. In the beginning, all ANNIE events will resemble those close to the wall of a much larger detector. In addressing these challenges, ANNIE provides an opportunity to understand the benefits of imaging photosensors. 

\subsubsection{Optimization of Timing-based Reconstruction Techniques}

Over the last several years, many novel approaches to WCh event reconstruction have been developed and applied to existing and proposed physics experiments. 

Pattern-of-light fitting techniques, such as those developed for the MiniBooNE collaboration~\cite{miniboone} and T2K~\cite{T2K2}, show promise as a way to maximally extract information from Water Cherenkov detectors. Another interesting approach uses Graphics Processing Units (GPUs) and ray-tracing algorithms to parallelize and quickly propagate photons through the large detector geometries~\cite{chroma}. However, the success of these techniques is limited by current PMT design. For example, in both the MiniBooNE and T2K reconstruction codes, the timing and charge likelihoods are factorized and calculated separately. This is connected to the fact that direct correlations between the positions and times of hits are lost in PMT electronics. With LAPPDs, it may be possible to implement these same approaches with a single likelihood based simultaneously on the positions and times of each hit. 
 
\begin{figure}
	\begin{center}
		\includegraphics[width=0.75\linewidth]{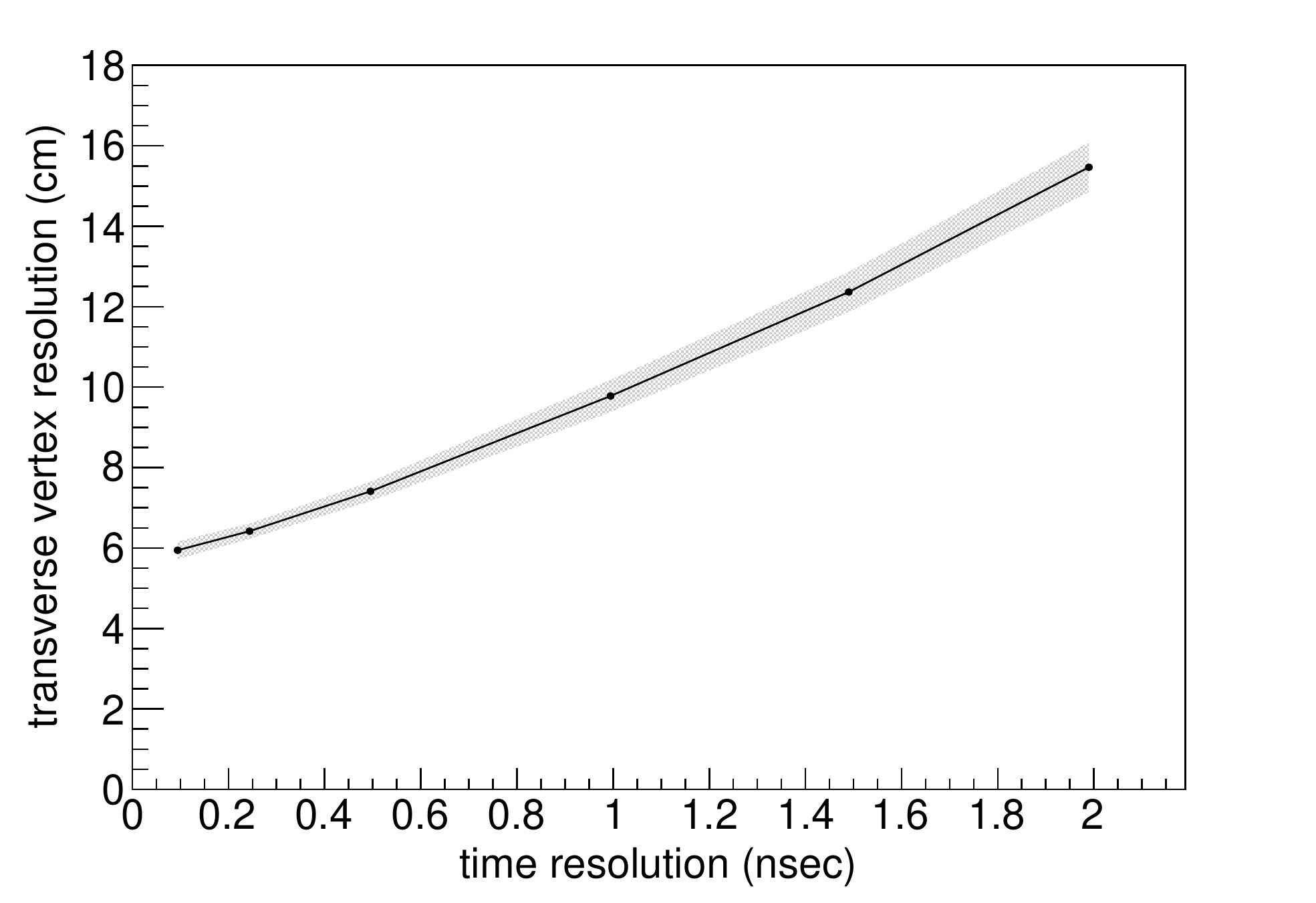}
	\end{center}
	\caption{Resolution of timing-based muon vertex fits in the direction transverse to the track direction, as a function of photosensor resolution. There is nearly a factor of 3 improvement as the photosensor resolution goes from 2 nsec to 100 psec.}
	\label{transverseres}
\end{figure}

Precision timing has shown promise in improving the capabilities of WCh detectors. Continuing work from a group based out of Iowa State University, the University of Chicago, and Argonne National Laboratory sees a factor of 3 improvement in muon vertex resolution for large, low-coverage detectors with 50 picosecond resolution, rather than a more typical 1.5-2 nanoseconds (Fig~\ref{transverseres})~\cite{fastneutrino}. 

\begin{figure}
	\begin{center}
		\includegraphics[width=0.65\linewidth]{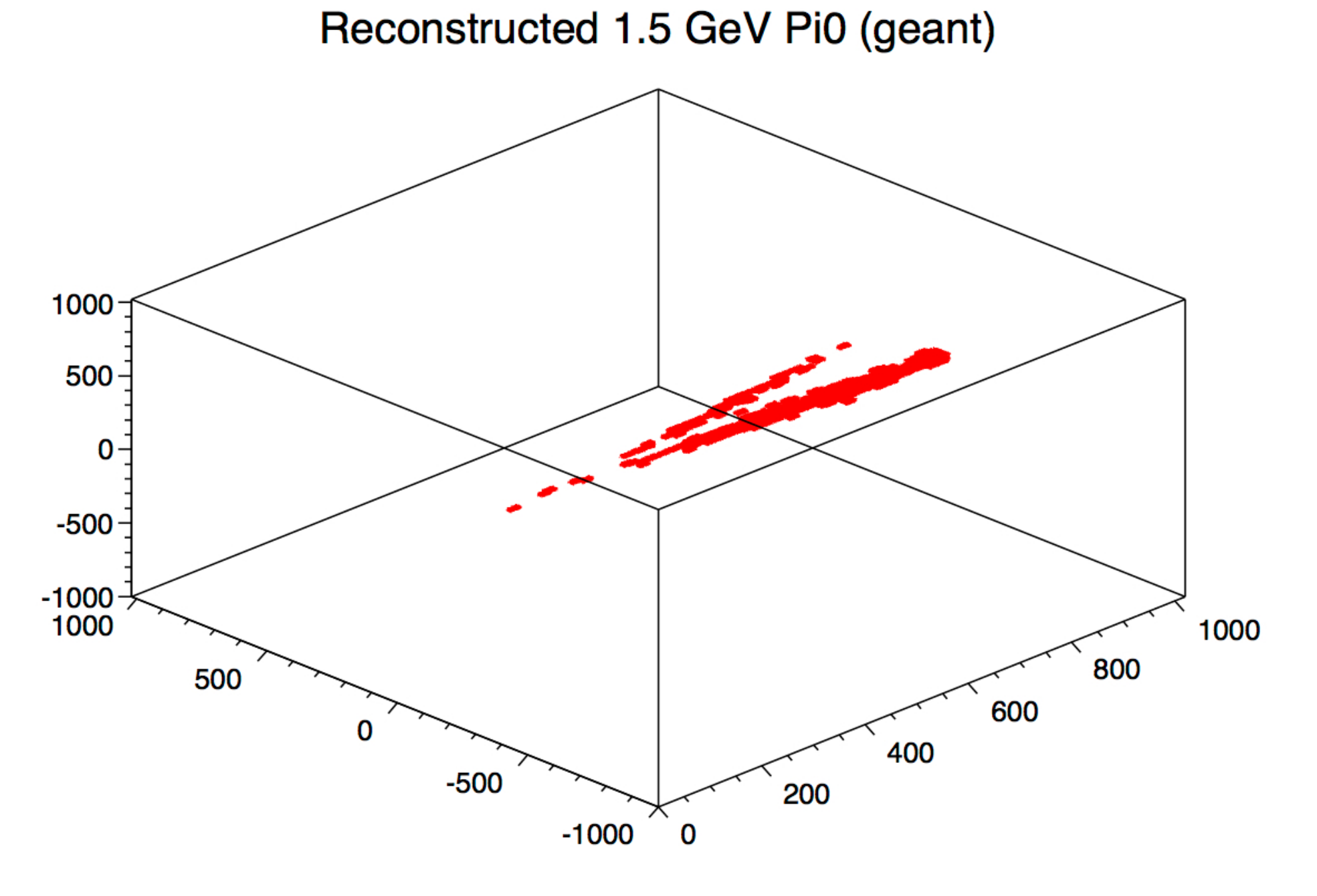}
	\end{center}
	\caption{Using a timing-based reconstruction algorithm to image the two gamma showers from the decay of a 1.5 GeV $\pi^0$.}
	\label{isochronfig}
\end{figure}

A causal Hough Transform could even be used to image tracks and EM showers from the  positions and times of detected photons, producing reconstructed event displays resembling those of Liquid Argonne detectors (Fig~\ref{isochronfig}). In fact, one can think of WCh detectors with fine time and spatial granularity as ``Optical Time Projection Chambers" (OTPCs) with the transit time of photons (instead of electron-ion pairs) used to reconstruct events~\cite{{OpticalTPC}}. Continued work on these techniques and, above all, validation in real data will be critical to the success and advancement of future optical tracking neutrino detectors. 

\section{Proposed Timescale}
\label{sec-schedule}

The data-taking phase of the ANNIE experiment is constrained by the operational lifetime of the booster neutrino beam line for the MicroBooNE experiment.  Also to maximize the impact of the R\&D on future water and liquid scintillator detectors the experiment the schedule here proposed should be started as soon as possible, ideally in FY2014. 

Year 1: Develop a complete MC of the ANNIE configuration to optimize water target and decide on required PMT and LAPPD coverage. Design the front-end system, trigger, and DAQ. Testing of electronics on laser test-bench at ANL.

Year 2: Construct water tank in the ANNIE Hall, install and test complete water system. Refurbish MRD. Decide on final design of FACC and install it after water target is in place. Demonstrate successful water-potting of LAPPDs. Operational test of LAPPDs on a small, known test beam or with radioactive samples on a small scale target. 

Year 3: Perform final commissioning and system integration in preparation for neutrino data-taking. Obtain first data-taking runs this year and begin data analysis. Install staged increase in LAPPD coverage.

Year 4: Continue additional data taking as necessary, including testing of various photodetector configurations and chemical enhancements to the water.

\section{Conclusion}
\label{sec-conclusion}

The ANNIE experiment provides an opportunity to make an important measurement of the final-state neutron abundance from neutrino interactions with water nuclei as a function of momentum transfer. This measurement will have a significant impact on a variety of future physics analyses, including a nearly factor of five (or more) improvement in the sensitivity of WCh proton decay searches provided by efficient neutron tagging. ANNIE also provides a low-cost opportunity to maintain technological diversity and Water Cherenkov expertise in the US neutrino program. ANNIE will represent a first demonstration detector for WCh reconstruction using the newly developed, high resolution LAPPD imaging sensors. Much development work has already been done by the LAPPD collaboration, including a working readout system with capabilities well matched to the needs of ANNIE. The experiment-specific development mainly involves implementation, some amount of customization, and long-term systems testing in a real physics context. 

We are still in the early stages of this experiment. Fermilab's support and interest will play a critical role in the success of ANNIE. Reserving the availability of the SciBooNE Hall for the duration of our measurement will ensure our access to the Booster Neutrino Beamline. Our effort will greatly benefit from support in the form of computer resources and simulations expertise. Finally, technical expertise and manpower in recommissioning the MRD and exploring the best ways to utilize ``ANNIE Hall" will also be useful.    


\end{document}